\documentclass[%
 reprint,
superscriptaddress,
 amsmath,
 amssymb,
 aps,
]{revtex4-2}
\usepackage[dvipsnames]{xcolor}
\usepackage{graphicx}
\usepackage{comment}
\usepackage{dcolumn}
\usepackage{bm}
\usepackage{hyperref}

\usepackage{xfrac}
\usepackage{siunitx}  
\usepackage[T1]{fontenc} 

\begin{document}

\preprint{APS/123-QED}

\title{Measuring nuclear spin qubits by qudit-enhanced spectroscopy in Silicon Carbide}

\author{Erik Hesselmeier}
\thanks{These two authors contributed equally}
 \affiliation{3rd Institute of Physics, IQST, and Research Centre SCoPE, University of Stuttgart, Stuttgart, Germany}

\author{Pierre Kuna}
\thanks{These two authors contributed equally}
 \affiliation{3rd Institute of Physics, IQST, and Research Centre SCoPE, University of Stuttgart, Stuttgart, Germany}
 

\author{Istv\'{a}n Tak\'{a}cs}
 \affiliation{Eötvös Loránd University, Egyetem tér 1-3, H-1053 Budapest, Hungary}
 \affiliation{MTA–ELTE Lend\"{u}let "Momentum" NewQubit Research Group, Pázmány Péter, Sétány 1/A, 1117 Budapest, Hungary}
 
\author{Viktor Iv\'{a}dy}
 \affiliation{Eötvös Loránd University, Egyetem tér 1-3, H-1053 Budapest, Hungary}
 \affiliation{MTA–ELTE Lend\"{u}let "Momentum" NewQubit Research Group, Pázmány Péter, Sétány 1/A, 1117 Budapest, Hungary}
 \affiliation{Department of Physics, Chemistry and Biology, Linköping
 	University, Linköping, Sweden}
  
\author{Wolfgang Knolle}
\affiliation{Department of Sensoric Surfaces and Functional Interfaces, Leibniz-Institute of Surface Engineering (IOM), Leipzig, Germany}
\author{Nguyen Tien Son}
 \author{Misagh Ghezellou}
\author{Jawad Ul-Hassan}
\affiliation{Department of Physics, Chemistry and Biology, Linköping
 	University, Linköping, Sweden}

\author{Durga Dasari}
 \affiliation{3rd Institute of Physics, IQST, and Research Centre SCoPE, University of Stuttgart, Stuttgart, Germany}

\author{Florian Kaiser}
  \affiliation{Materials Research and Technology (MRT) Department, Luxembourg Institute of Science and Technology (LIST), 4422 Belvaux, Luxembourg}
  \affiliation{University of Luxembourg, 41 rue du Brill, L-4422 Belvaux, Luxembourg}

\author{Vadim Vorobyov}
\email[]{v.vorobyov@pi3.uni-stuttgart.de}
 \affiliation{3rd Institute of Physics, IQST, and Research Centre SCoPE, University of Stuttgart, Stuttgart, Germany}  

\author{J\"org Wrachtrup}
 \affiliation{3rd Institute of Physics, IQST, and Research Centre SCoPE, University of Stuttgart, Stuttgart, Germany}
 \affiliation{Max Planck Institute for solid state physics, Stuttgart, Germany}


\begin{abstract}
Nuclear spins with hyperfine coupling to single electron spins are highly valuable quantum bits. 
In this work we probe and characterise the particularly rich nuclear spin environment around single silicon vacancy color-centers (V2) in 4H-SiC.
By using the electron spin-3/2 qudit as a 4 level sensor, we identify several groups of $^{29}$Si and $^{13}$C nuclear spins through their hyperfine interaction. 
We extract the major components of their hyperfine coupling via optical detected nuclear resonance, 
and assign them to shell groups in the crystal via the DFT simulations. 
We utilise the ground state level anti-crossing of the electron spin for dynamic nuclear polarization and achieve a nuclear spin polarization of up to $98\pm6\,\%$.
We show that this scheme can be used to detect the nuclear magnetic resonance signal of individual spins and demonstrate their coherent control. 
Our work provides a detailed set of parameters for future use of SiC as a multi-qubit memory and quantum computing platform.
\end{abstract}

\maketitle
\section{Introduction}
Nuclear spins are outstandingly robust quantum bits. Their spin coherence times can extend over hours and spin relaxation can be prolonged to even longer time scales \cite{Zhong_2015}. As a result, they rank among the most precisely controlled quantum systems. They have been used to demonstrate multiparticle entanglement \cite{neumann2008multipartite, childress2006coherent, bradley2019ten}, quantum non-demolition readout \cite{neumann2010single, maurer2012room, dreau2013single}, quantum memories robust for optical cycles for quantum repeater applications \cite{bradley2022robust,vorobyov2023transition} as well as quantum registers to demonstrate quantum error correction \cite{waldherr2014quantum, Abobeih_2022} or quantum simulations \cite{randall2021many}. 
These properties make them an excellent platform for emerging quantum technologies, specifically for quantum networks with a current record of around 50 addressable nuclear quantum bits \cite{van2023mapping}.

However, most individual nuclear spins are hard to probe, due to their weak interaction with the central spin and applied RF fields.
This is different when they are close to the single electron spin, which drastically increases the interaction strength. 
Such single nuclear spins have been detected and used in various host materials \cite{WRACHTRUP1997179,M_dzik_2022,Pla:2014aa,Atat_re_2018, Gao_2022}. 
This also comprises $^{13}$C as well as $^{29}$Si nuclei in silicon carbide (SiC) \cite{Awschalom_2018} where the latter material is an example of a system with a diatomic lattice where the isotope composition can be precisely controlled. 

\begin{figure*}
	\centering
	\includegraphics[width=\textwidth]{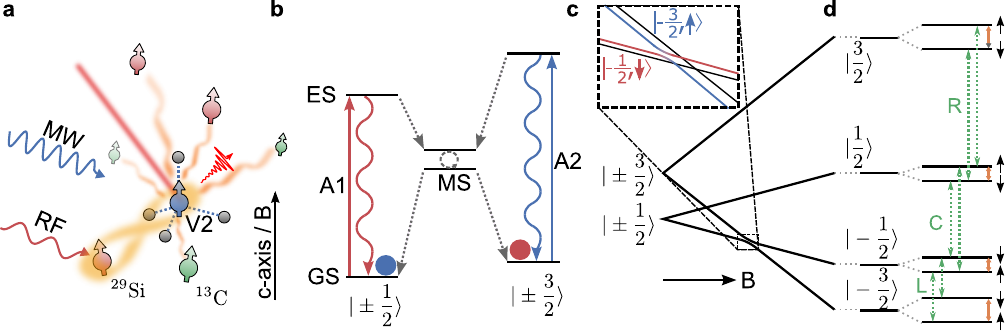}
	\caption{\textbf{(a)} The silicon vacancy electron spin is coupled to various nuclear spins by strong hyperfine interaction. Its state can be manipulated via optical and microwave transitions. Nearby nuclear spin isotopes can be used as qubits and manipulated via RF pulses. \textbf{(b)} Energy level diagram of the V2 center in the absence of a magnetic field. At cryogenic temperatures, A1 and A2 are individually addressable spin-selective transitions between the electronic ground (GS) and excited state (ES) allowing for readout of the spin state. Decay processes from the ES to the metastable state (MS) lead to polarization of the electron spin. \textbf{(c)} Energy levels in the ground state. The inset shows the ground state level anticrossing between the $m_s=-1/2$ and $m_s=-3/2$ sublevel. \textbf{(d)} Green arrows show allowed ($\Delta m_s=\pm1$) spin transitions, where L, C, and R denote left, center, and right, according to the position in the ODMR spectra. Orange transitions are nuclear single-quantum transitions.}
	\label{fig1}
\end{figure*}

\textcolor{black}{The V2 center in 4H-SiC} has shown particularly favorable properties, motivating its use as efficient spin photon interface \cite{banks2019resonant}\textcolor{black}{, e.g. preserved optical coherence at temperatures up to 20\,K \cite{Udvarhelyi2020} and tolerable spectral diffusion in nanophotonics \cite{babin2022fabrication}. Its electronic spin properties and inter-system crossings have been investigated in detail \cite{liu2023silicon} and two-photon interference has been shown \cite{Morioka2020}.}
What is missing so far is an extensive characterisation of the nuclear spin environment of the V2 center. 
So far ensemble electron spin resonance spectroscopy and density functional theory (DFT) calculations have been used \cite{janzen_silicon_2009,ivady2017identification,tarasenko_spin_2018} to obtain the hyperfine tensor of a few types of nuclear spins \cite{soltamov2021electron}.

In this work, we carry out an extensive study on the hyperfine coupling of various nuclei surrounding single V2 centers and achieve coherent control in the vicinity of the ground-state level anticrossing (GSLAC).
We show that under these conditions, nuclear spins can be efficiently optically polarized and, in contrast to the case of NV system described in \cite{sangtawesin2016hyperfine}, possess hyperfine enhancement of nuclear g-factor of various strengths in all four electron subdomains allowing it to be efficiently probed. 
The richness of a spin-3/2 system as a four-level qudit quantum sensor allows us to obtain major terms of the hyperfine tensor accurately even when measured at a single value of the magnetic field. 
In theory, spin-1 (NV-like) systems also allow for the resolution of the major components of the hyperfine tensor near the GSLAC, however, its experimental realisation is more challenging \cite{Auzinsh2019GSLAC}.
In addition, we carried out extensive DFT calculations for the V2 center to obtain finite-size effect-free theoretical hyperfine data.
By comparison with the calculations we can assign the investigated nuclear spins to distinct groups of lattice sites in the crystal.
Finally, we probe the coherence properties of various nuclear spins in all subdomains.
Our results provide a necessary step on the way toward using nuclear spins around the V2 center in quantum information and quantum sensing applications. 

\section{Results}
The V2 center is formed by a missing silicon atom at a cubic lattice site substituted by an electron, giving rise to a spin-3/2 system as depicted in Fig. \ref{fig1}a. 
Its optical dipole is oriented along the crystal c-axis and emits light in the zero-phonon line at \SI{916.5}{\nano\meter}. 
The V2 center exhibits two optical transitions, A1 and A2, which are separated by $\sim$\SI{1}{\giga\hertz} \cite{nagy2019high}. 
The A1 (A2)-transition belongs to the $m_s=\pm1/2$ $(\pm3/2)$ electron spin submanifold (Fig. \ref{fig1}b) which can be used for readout of a specific spin sublevel as well as its initialization. 
The Hamiltonian of the ground state, when the $B$-field is aligned along the c-axis ($B_z$) of the crystal, reads:
\begin{equation}
H = D \left(S_z^2 + \frac{1}{3}S\left( S+1\right)\right) + \gamma B_z S_z + \vec{S} \cdot \textbf{A} \cdot \vec{I} + \gamma_n B_z I_z
\end{equation}
The ground state zero-field splitting (ZFS) $D=\SI{35}{\mega\hertz}$ lifts the degeneracy between the $m_s = \pm1/2$ and $\pm3/2$ sublevels. Further degeneracy of the sublevels can be removed by applying a magnetic field (Fig.~\ref{fig1}c). 
$\textbf{A}$ is the hyperfine tensor. 
At magnetic fields yielding an electron Zeeman interaction larger than $\textbf{A}$, the hyperfine interaction can be treated as a zero-order perturbation.
In this approximation only electron spin transitions with $\Delta m_s=1$ are allowed. This results in three hyperfine doublets, labeled as L (left), C (central), and R (right), (Fig.~\ref{fig1}d). 
The precise measurement of all components of the hyperfine tensor allows us to refine the spin wavefunction of the electron spin and extract the location of the nuclear spin. 
The most precise way to measure the components of the $\textbf{A}$ tensor is to rotate the external magnetic field. 
However, rotation and precise determination of the magnetic field components $B_{x,y,z}$ is challenging in most single spin experiments. 

\begin{figure*}
	\centering
	\includegraphics[width=\textwidth]{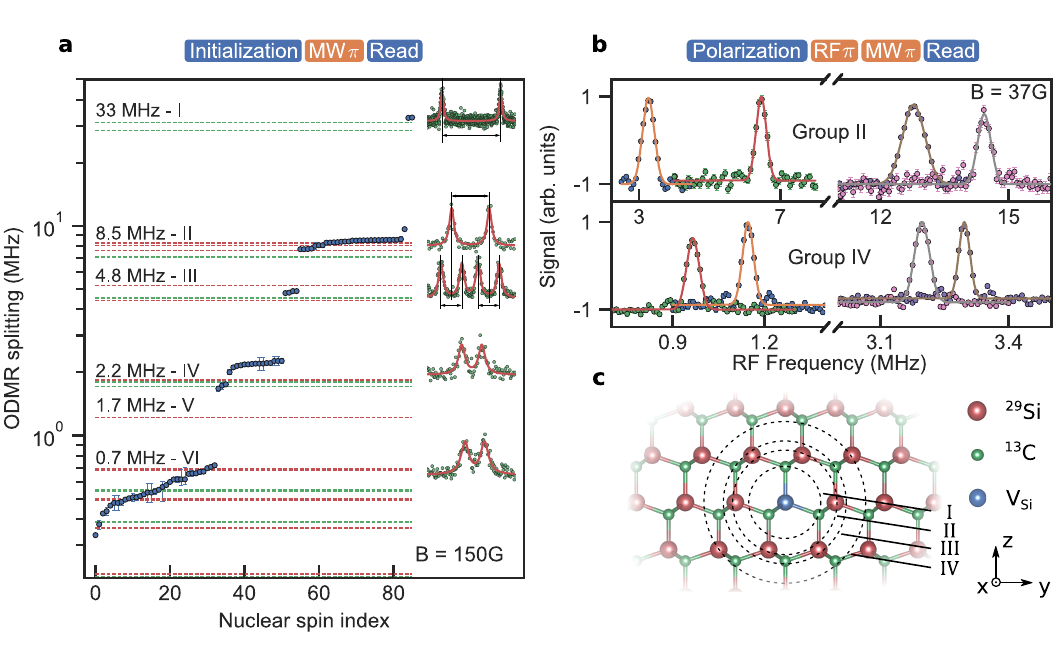}
		\caption{\textbf{(a)} ODMR Splittings of $m_s =-1/2$ to $m_s=-3/2$ transition at $B=150$\,G. Red (green) dotted lines indicate ODMR-splittings due to $^{29}$Si ($^{13}$C)-isotopes. The discrete steps correspond to the distinct lattice sites around the V2 center. \textbf{(b)} Electron-nuclear double resonance measurements of group II and IV nuclear spins. \textbf{(c)} Crystal structure of 4H-SiC. Dotted circles indicate shells of lattice sites with similar ODMR splitting.}
	\label{fig2}
\end{figure*}

Here, we use a different method to determine the components of $\textbf{A}$.
In the region of the GSLAC, the exact frequency of the electron spin and nuclear spin transitions sensitively depends on the components of $\textbf{A}$.
Moreover all electron and nuclear single quantum (SQ) spin transitions can be directly measured through ODMR and optically detected nuclear magnetic resonance (ODNMR).
In the first step, we record ODMR spectra at 150\,G and extract the splitting of the electron transitions. To initialize and read out the electronic spin state, we use the sequence as described in Ref.~\cite{nagy2019high}. We carry out these measurements for more than 70 defects and extract the splittings from the ODMR spectra (Fig.~\ref{fig2}a). 

In our statistics, we observe multiple discrete values in the splittings and correlate them with the various lattice sites. Note, that 8\,MHz and 2\,MHz splittings are observed more often than 30\,MHz and 4\,MHz splittings. We attribute this to the approx.\ 4 times higher abundance of $^{29}$Si than of $^{13}$C isotopes. 
Quite strikingly the simulated values for the hyperfine coupling (shown by the dashed lines) agree very well with measurements in Fig.~\ref{fig2}a. To obtain the position of the hyperfine transition from DFT data\footnote{The hyperfine data is available online at ...}, we calculated the $\textbf{A}$ tensor from electron spin density, included the experimental $B$-field magnitude and orientation, and derived the anticipated position of the resonance line. The stepwise change in the ODMR splitting in the experiment and theory indicates the shell-like arrangement of nuclear spins around the defect center.
With the hyperfine tensor information at hand, we can attribute measured couplings to specific nuclear spin locations in the lattice which we denote as I-VI.  Even small sub-steps, like the one observed in group II nuclear spins, are accurately reproduced.
Additionally, it allows to select among the various nuclear spins those which are preferable for hyperpolarization, quantum memory or quantum gate operations. 

To experimentally determine the components of the hyperfine coupling we now proceed by repeating the ODMR measurements of all electron transitions close to the GSLAC. 
We select V2 centers that show strong coupling to nuclear spins of different groups and precisely measure the transition frequencies.
With the assumption that strongly coupled spins are dominated by the Fermi contact, $A_{zy} = 0$, $A_{xy} =0$ and $A_{xx}\approx A_{yy}$. In second-order perturbation, the transition frequencies between the states $|-1/2, \downarrow \rangle$ and $|-3/2,\downarrow \rangle$ are
\begin{equation}
\Delta E = \gamma_e B -2D - \frac{1}{2}A_{zz} - \frac{A_{zx}^2}{4A_{zz}} + \frac{3}{16} \frac{(A_{xx} + A_{yy})^2}{\gamma_e B -2D + A_{zz}}.
\end{equation}
It shows, that the ESR transitions thus depend on various tensor elements of $\textbf{A}$ \textcolor{black}{(see SM: IV for equations of all 6 allowed transitions)}. 

To infer the hyperfine coupling elements of the nuclear spins more precisely we combine qudit spectroscopy with a direct measurement of the nuclear qubit transition frequency in all electron spin subdomains. 
After charge initialization (not shown), electron and nuclear spins are initialized via dynamic nuclear polarization due to flip-flop interactions (DNP, see below).
We then sweep radio-frequency excitation through the nuclear magnetic resonance transition in each electron spin sublevel. 
When the frequency matches the nuclear resonance, a nuclear selective narrowband MW $\pi$ pulse results in a concomitant change in fluorescence. 
To measure nuclear transitions in the $m_s=+1/2$ and $m_s=+3/2$ state, the system is also prepared in $m_s=-1/2$ state and cascaded into the  $+1/2$ and $+3/2$-states by MW  $\pi$-pulses on the C and R transition, respectively.
The NMR transitions for the most common spins, i.e. group II and IV, are shown in Fig.~\ref{fig2}b.

For a detailed analysis, we perform a numerical fitting of the system Hamiltonian to all electron and nuclear spin transition frequencies and thus yield the hyperfine interaction tensor (see SM: I). 
In the specific case of the nuclear spin from group II, shown in Fig.~\ref{fig2}b, the fit gave a magnetic field of $B=36.83\pm0.01$\,G, a zero-field splitting of $D=34.89\pm0.01$\,MHz and the following coupling matrix entries between the electron and nuclear spin (see Table~\ref{tab1}).

\begin{table}[b]
\begin{center}
	\begin{tabular}{|c|c|c|c|c|c|c|c|}
	\hline 
	G& res. &$A_{zz}$  &$A_{xx}$ &$A_{yy}$ &$A_{iso}$ & $T$ \\ [0.5ex] 
	\hline
     Si$_{II}$& 0.163 &8.660(3) &9.00(1) &9.03(1) & 8.910(6) & -0.130(4) \\ [1ex] 
	C$_{III}$& 0.38 & 4.9(4) &6.4(6) &6.4(5) &5.96000(2) &-0.4900(3)\\ [1ex]
	 Si$_{IV}$& 0.14&-2.200(3)& -2.7(2)&-2.6(2)& -2.48 &0.14\\ [1ex] 
	C$_{VI}$& NA & 0.42  & <1.35 & <1.35 &NA& NA\\ [1ex] 
	\hline
	\end{tabular}
	\caption{\label{tab1} The fitted values for the couplings (units are in MHz)}
\end{center}
\end{table}

\begin{figure*}[t]
	\centering
	\includegraphics[width=\textwidth]{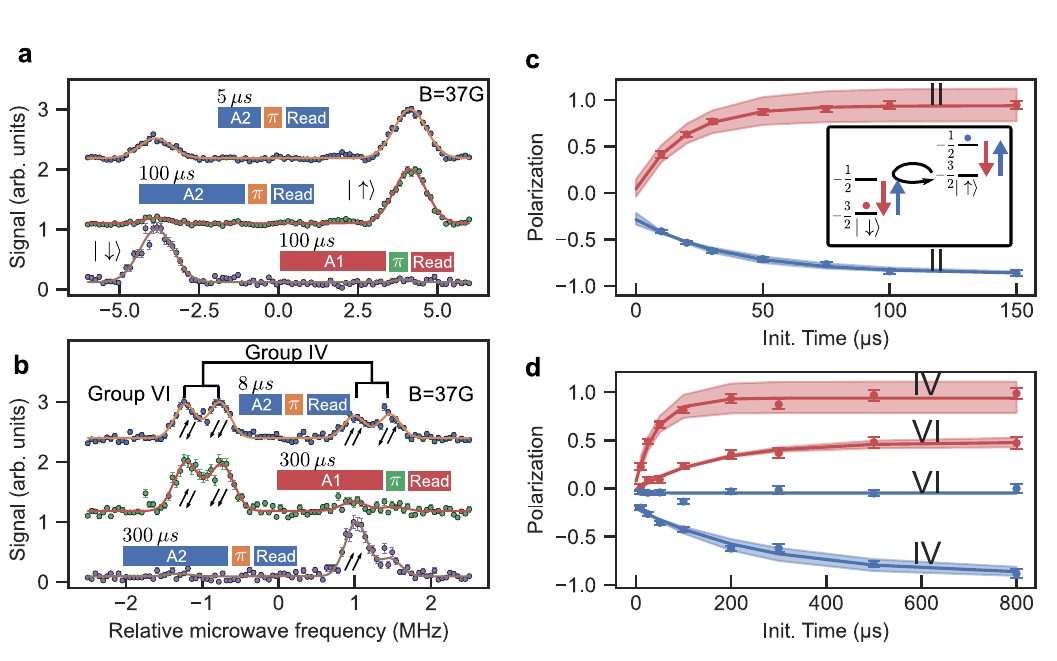}
	\caption{\textbf{(a)} ODMR measurements near the GSLAC after initalization times of 5 and $100\,\mu$s, respectively. Due to interactions between the electron and a group II nuclear spin, polarization can be transferred from the electron to the nuclear spin.  \textbf{(b)} ODMR measurements of a V2 center strongly coupled to two nuclear spins.\textbf{(c)} Achieved nuclear polarization after various initialization times. Inset shows the dynamics, which lead to polarization of the nuclear spin. \textbf{(d)} Nuclear spin polarization of group IV and VI nuclear spins for various initialization times fitted with exponential curves $\sim 1-\exp(-t/T)$ with characteristic time $T = \SI{44} {\mu s}$ for curve IV and $T = \SI{162}{\mu s}$ for VI}
	\label{fig3}
\end{figure*}

Next, we use the information obtained on $\textbf{A}$ to investigate the polarization of nuclear spins.
\textcolor{black}{The V2 center electron spin can be efficiently polarized via optical pumping \cite{nagy2019high, liu2023silicon}. In general, this is not the case for nuclear spins, as they are insensitive to optical excitation.}
Quantum gates can be used for efficient transfer of polarization by frequency selective CNOT gates \cite{bourassa2020entanglement, Pagliero_2014}. 
However, this requires a high-quality gate performance which is particularly difficult in dense nuclear spin environments. 
Earlier it has been shown that level anticrossings can be used to efficiently polarize nuclear spins with strong hyperfine coupling to the electron spin \cite{Jacques2009, Smeltzer2011, Falk2015working}.  

Here, we use V2 center's intersystem crossing and spin-selective optical excitation of the electron spin to achieve high nuclear spin polarization levels. 
By carefully adjusting the $B$ field, we tune the defects’ spin sublevels close to the level anticrossing points, such that the hyperfine interaction hybridizes states with the same total spin number $m_s + m_I$ allowing zero quantum transitions ($\Delta m = 0$). 
Due to the qudit nature of the electron spin, level anticrossings in the ground state occur at various fields. Here, we focus on the transition between $m_s = -3/2$ and $m_s=-1/2$ at approximately \SI{37}{G} slightly above the GSLAC of \SI{25}{G}.
A detailed theoretical model of the dynamics under GSLAC conditions can be found in \cite{Ivady2015}. 
We thus restrict ourselves to a qualitative explanation of the polarization process. 

At the 25\,G GSLAC condition the transition energy between $m_s=-3/2$ and $m_s=-1/2$ vanishes (see Fig \ref{fig1}c). This leads to strong mixing between  $|-1/2,\uparrow\rangle$ and $|-3/2,\downarrow\rangle$ due the nonsecular terms in the hyperfine part of the Hamiltonian, namely $V = A_\perp (S_x I_x + S_y I_y) = A_\perp (S^- I^+ + S^+ I^-)/2$. 
This flip-flop interaction can be used to transfer polarization from the electron to the nuclear spin. 
The polarization rate is proportional to $A_\perp$, mainly dependent on the dominant $A_{xx}$ and $A_{yy}$ terms, and the rate at which the electron is reinitialized. 
Under optical saturation, it typically is $\gamma \sim A_\perp^2/(A_\perp^2+\delta^2)$ which sets the polarization and decay rates \cite{neumann2010single}.

In ODMR measurements on the $m_s = -3/2$  to $m_s = -1/2$ transitions we observe robust nuclear hyperpolarization as revealed by the hyperfine resolved ODMR spectrum (Fig.~\ref{fig3}a). Without polarization, both nuclear spin states appear in the spectrum.  
Under A1 excitation the mixing between $|-3/2,\downarrow\rangle \leftrightarrow|-1/2,\uparrow\rangle$ polarizes the system into the one state which is not excited by the laser and is not efficiently coupled by spin flip-flop processes to any other level, namely the $|-3/2,\uparrow\rangle$ state.  
By changing to A2 excitation, we polarize the system into state $|-1/2,\downarrow\rangle$ as shown in Fig.~\ref{fig3}a. 
By comparing the peak heights in the ODMR at various initialisation times, we estimate the population and verify a polarization of  $94.6\pm4.0\,\%$ and $86.5\pm3.3\,\%$ of initialising into the state $|-3/2, \downarrow\rangle$ and $|-1/2 \uparrow \rangle$ under A1 and A2 excitation respectively. 
We note however, that this is the polarization within the $m_s\in\{-3/2, -1/2\}$ manifold. 
The tomography of all electron levels as a function of initialisation time, which was done via the Rabi nutation contrast (see SM: II),  shows the initialisation values of 88 $\pm$ 5 \% 78 $\pm$ 10\% respectively.

Fig.~\ref{fig3}c shows a case of a V2 defect with two coupled nuclear spins belonging to two different groups (IV and VI) with hyperfine coupling parameters ($A_{zz}=2.2$\,MHz and 0.6\,MHz). 
Due to their asimilar hyperfine coupling they polarize with a contrasting rate (see Fig. \ref{fig3}d) and hence can be selectively polarized (see Fig.\ref{fig3}c.)
In general, the polarization in $m_s=-1/2$ is smaller than for $m_s=-3/2$, due to leakage via the detuned flip-flop interactions between $|-1/2,\downarrow\rangle$ and $|+1/2,\uparrow\rangle$.
Here we estimate the lower bound of the initialisation fidelity for the weaker coupled nuclear spins, by simply analysing the relative amplitude in the ODMR split peaks. 
The polarizations achieved (98 \%, 88\% for II, and 47\% for IV) here are sufficiently high to be used for characterisation purposes, and for prepolarization, for subsequent algorithmic or measurement-based initialization.


\begin{figure*}[ht]
\centering
	\includegraphics[width=\textwidth]{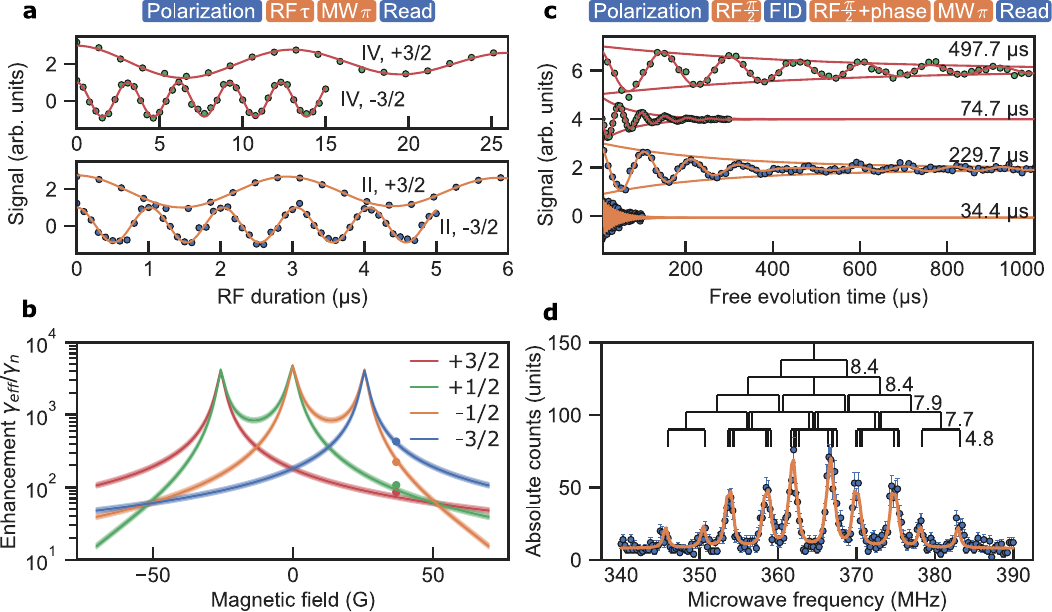}
	\caption{\textbf{(a)} Rabi drive of group II and IV nuclear spins in $\pm3/2-$electron spin sublevels. The same RF power is used in all measurements. Errors are smaller than data points.\textbf{(b)} Analytically calculated hyperfine enhancement of nuclear g-factor of a group IV nuclear spin coupled to a V2 center. Dots represent measured values. \textbf{(c)} Corresponding Ramsey fringes in (+3/2, -3/2) - electron spin sublevels. $T_2^\ast$ is drastically reduced in $m_s=-3/2$. Errors are smaller than data points. \textbf{(d)} ODMR of V2 center simultaneously coupled to five nuclear spins with 8.4,8.4,7.9,7.7,4.8 MHz splittings at 150\,G. Tree structure indicates the various resulting splittings due to each nuclear spin.}
	\label{fig4}
\end{figure*}

We further measure Rabi oscillations of the nuclear spins. 
Here, instead of sweeping the frequency of the radiofrequency drive we tune it into resonance and measure the response of the nuclear spin for increasing pulse duration. The distinct hyperfine coupling as well as strong spin polarization allows us to efficiently address single nuclear spins and coherently control them. Fig.~\ref{fig4}a shows Rabi oscillations of group II and IV nuclei in the $m_s=\pm3/2$ electron spin states. 
We recall, that since all avoided crossings between the levels appear at rather small magnetic fields ($-25$ G, $0$ G, $+25$ G) they are of the same order as the hyperfine interaction.
This leads to non negligible hyperfine enhancement of various strength in all four electron spin sublevels. 
In general, the closer it is to the GSLAC, the larger is the hybridization of nuclear and electron spin, leading to effective hyperfine enhanced gyromagnetic ratios of the nuclear spin \cite{Chen2015, sangtawesin2016hyperfine}. 
The enhancement $\alpha$ of the nuclear gyromagnetic ratio in $m_s = -3/2$ reads
\begin{align}
    \alpha_{-\frac{3}{2}} &= -\sqrt{3}\frac{\gamma_e}{\gamma_n}\sin(\vartheta^L)+\cos(\vartheta^L)\\
    \tan(2\vartheta^L) &= \frac{-\frac{\sqrt{3}}{2}(A_{xx}+A_{yy})}{-A_{zz}+B_z(\gamma_n-\gamma_e)+2D},
\end{align}
with $\gamma_e$ $(\gamma_n)$ being the electron (nuclear) gyromagnetic ratio. 
Our experimental observations match well with the theoretical model (see Fig. \ref{fig4}b).
This enhancement leads to fast nuclear Rabi nutation even at low RF driving fields (see Fig. \ref{fig4}b, detailed description in SM: V), with strongest enhancement in $m_s=-3/2$ states, and least in $m_s=+3/2$.
Note, that while $\pm3/2$ states have a single enhancement peak at corresponding GSLAC B-field values, the $m_s=-1/2$ ($m_s=1/2$) sublevels show two peaks at 0 and at $-25(25)$ G, resulting in increased and broader enhancement across the B-field. 

For estimation of the coherence properties, we perform Ramsey interferometry on the nuclear spins in various electron sublevels. 
As previously noted \cite{sangtawesin2016hyperfine}, the enhancement of the gyromagnetic ratio also leads to a shortening of the coherence time, due to increased sensitivity to the magnetic noise. 
In the electron $m_s=+3/2$ subdomain, where the hyperfine enhancement is the smallest, Ramsey fringes decay on a time scale up to $T_{2, +3/2}^\ast\approx \SI{500}{\micro\second}$ for a group IV isotope (Fig. \ref{fig4}c). 
A possible cause for the decay is electron spin relaxation. 
We note, however, that the $T_1$ time of the electron spin is longer than 10\,ms. 
Hence we conclude, that coupling to nuclear or electron spins in the environment limits the nuclear $T_2^\ast$ as coupling strength on the order of kHz is expected for the nuclear spin concentration of our sample with elevated gyromagnetic factor. 
Indeed, if we measure the Ramsey fringe decay in other electron spin states, which are closer to the GSLAC and thus have a larger electron-nuclear hybridization, the decay times are reduced to $T_{2, -3/2}^\ast \approx \SI{74}{\micro\second}$.
Also, coherence times seem to decrease the stronger the hyperfine interaction with the electron is. 
This is due to a higher level of hybridisation between electron and nuclear states for larger values of the $\textbf{A}$ tensor in the GSLAC vicinity. 

\section{Discussions}
Owing to its electron quartet ground state and $^{29}$Si as well as $^{13}$C nuclear spins, the V2 center shows a very rich variety of potential nuclear spin qubits and control options via their hyperfine coupling to the electron spin. 
In combination with the long electron spin coherence times and excellent electron spin initialization nuclear spins can be initialized with high fidelity and selectivity. 

Our work represents a significant step forward compared to previous experiments in which weak coupling between single silicon vacancy centers and 1-2 nuclear spins was observed. 
We provided an extensive screening of dozens of nuclear spins occurring in the first six shells around the V2 center. We showed that the S=3/2 qudit nature of the V2 center allows to \textit{accurately infer the entire hyperfine tensors} using only one set of experimental parameters at constant $B$ field. 
We used this knowledge to polarize and control nuclear spin qubits with high fidelities reaching 98\%. 
Our SiC sample had a natural isotope abundance (4.7\,\% $^{29}$Si and 1.1\,\% $^{13}$C), which resulted in a high abundance of strongly coupled nuclear spins, e.g., we observed up to five strongly coupled spins depicted in Fig. \ref{fig4}d. 
These nuclear spins are excellent candidates for achieving single-shot electron spin readout via a nuclear spin ancilla qubit \cite{Jiang2009}. Additionally, quantum sensing applications can benefit from strongly coupled nuclear spins, e.g., via error correction \cite{unden2016quantum} or in-situ quantum Fourier transformation \cite{vorobyov2021quantum} for extending the dynamic range of the sensor. 
Also, individual nuclear spins have been proven to be very instrumental in achieving high-resolution nanoscale nuclear magnetic resonance \cite{aslam2017nanoscale}. Here, nuclear spins serve as room-temperature quantum memory to increase correlation times. So far only a single nuclear spin in diamond has been used. With their larger abundance in SiC, this can be extended to the usage of multiple nuclear qubits for the construction of e.g. decoherence-free subspaces among the strongly coupled nuclear spin owing to symmetric atomic site placements and similar hyperfine tensor.

Many quantum information applications from the memory, communication and computation branches are likely to require isotopically engineered SiC crystals with an abundance on the order of $\approx 1\%$ to reduce the amount of strongly coupled nuclear spins \cite{bourassa2020entanglement,Parthasarathy2023}. This provides unperturbed access to weakly coupled nuclear spins, which usually results in drastically longer coherence times \cite{Petersen2016}. Additionally, weakly coupled nuclear spins show a strongly reduced vulnerability against optical excitation-relaxation of the central color center \cite{Bradley_2022, vorobyov2023transition}.
Further improvements of nuclear spin coherence times can be achieved by operation in increased magnetic fields \cite{Yang2014}, and nuclear spin gate fidelities can be improved using adapted control techniques \cite{Bradley2019, van2023mapping}.

In summary, our results qualify the V2 center in 4H-SiC as a highly attractive system in the framework of nuclear spin qubit-enhanced quantum applications across all fields. We expect the full potential of the system to be unleashed with the availability of isotopically engineered SiC material.



\section*{Acknowledgments}
F.K. and J.W. acknowledge support from the European Commission for the Quantum Technology Flagship project QIA (Grant agreements No. 101080128, and 101102140).
P.K., F.K., J.U.H., and J.W. acknowledge support from the European Commission through the QuantERA project InQuRe (Grant agreements No. 731473, and 101017733).
P.K., F.K. and J.W. acknowledge the German ministry of education and research for the project InQuRe (BMBF, Grant agreement No. 16KIS1639K).
F.K. and J.W. further acknowledge the German ministry of education and research for the project QR.X (BMBF, Grant agreement No. 16KISQ013), while J.W. also acknowledges support for the project Spinning (BMBF, Grant agreement No. 13N16219).
F.K. and J.W. additionally acknowledge the Baden-Württemberg Stiftung for the project SPOC (Grant agreement No. QT-6).
J.U.H. and V.I. further acknowledge support from the Swedish Research Council under VR Grant No. 2020-05444 and Knut and Alice Wallenberg Foundation (Grant No. KAW 2018.0071).
J.U.H. acknowledges The Swedish Research Council VR under VR Grant No. 2020-05444.
V.I. and I.T. was supported by the National Research, Development, and Innovation Office of Hungary within the Quantum Information National Laboratory of Hungary (Grant No. 2022-2.1.1-NL-2022-00004) and within grant FK 145395. The calculations were performed on resources provided by the Swedish National Infrastructure for Computing (SNIC) at the National Supercomputer Centre (NSC).

\end{document}